\documentclass[journal=dce]{CUP-JNL-DTM}

\usepackage{amsmath}
\usepackage{amssymb}
\usepackage{graphicx}
\usepackage{booktabs}
\usepackage[T1]{fontenc}

\jname{Data-Centric Engineering}
\jyear{2026}
\jvol{xx}
\jissue{xx}

\begin{document}

\title[Null-Corrected Conditional Coverage]{Null-Corrected Conditional Coverage
in Materials Property Prediction: A Five-Dataset Replication}

\author[1]{Irene Boruah}

\address[1]{Department of Mechanical Engineering, Dibrugarh University Institute
of Engineering and Technology, Dibrugarh, 786004, Assam, India.
Corresponding author: ireneboruah2004@gmail.com}

\keywords{conformal prediction; conditional coverage; uncertainty
quantification; replication; materials property prediction}

\abstract{Uncertainty quantification for materials property prediction increasingly
reports conditional coverage, coverage broken out by strata of the predicted
value, alongside marginal coverage. The worst-bin statistics used for this are
downward-biased in finite samples, a defect known in the conformal prediction
literature but not generally corrected for in applied materials work, where
per-bin sample sizes are smallest and the bias is largest. We replicate a
single-dataset conditional coverage study across five regression datasets
($n=437$ to $21{,}263$), seven interval methods and 50 random splits each, the seventh a level-matched ablation run on the two smallest datasets, correcting the worst-bin
statistic against its expected value under exact conditional validity. The null
is simulated from realized per-bin test counts and referenced to each method's
own marginal coverage, separating marginal miscalibration from uneven coverage
across bins. On the smallest dataset it runs from 0.061 to 0.096, between a
third and two thirds of the shortfalls actually reported, against 0.009 to
0.011 on the two largest. Split conformal prediction's failure survives the
correction on four of five datasets; two conclusions of the antecedent study do
not. Conformalized quantile regression, characterized there as a negative case,
fails on one dataset only, where its correction term degenerates to zero.
Cross-fitted normalized conformal prediction, grouped there with the successful
methods, fails on three of five. Group-conditional (Mondrian) conformal prediction stays below our significance criterion on every dataset, at exceedance rates of 0.10 to 0.24, yet exceeds the null on three of five under an octile refinement of the variable it calibrates on. A level-matched ablation locates that uniformity in group-conditional
calibration rather than in the elevated effective quantile level, and removes
Mondrian's apparent width cost against split conformal prediction entirely. We
recommend that conditional coverage evaluations report the null distribution of
whatever worst-bin statistic is used, the method's realized marginal coverage,
and the resolution of the evaluation partition relative to any partition the
method calibrates on.}

\maketitle

\section*{Impact Statement}

Engineers who use machine learning to set design allowables need prediction
intervals that hold where the decision is made, not only on average. Materials informatics now checks this by reporting coverage within strata of the predicted property and flagging those that fall short. This paper
shows that the statistic used for that check is biased against the method being
tested, and that on datasets of the size common here it can account for most of the reported failure. Correcting it across five datasets
and seven methods overturns two conclusions of a companion manuscript under review about which methods are
safe to use, and bounds how far a group-conditional guarantee transfers to
groupings finer than the one it calibrates on.

\section{Introduction}

Conformal prediction gives finite-sample marginal coverage under
exchangeability, and it is established that marginal coverage does not imply
conditional coverage: distribution-free conditional validity is impossible
without further assumptions \citep{vovk2012,barber2021limits}, and
constant-width intervals undercover wherever residual scale exceeds average.
Applied work in materials property prediction has responded by adopting
locally adaptive interval methods and by reporting coverage stratified by
predicted value.

The diagnostic underlying that practice is weaker than its use suggests.
Reporting the minimum coverage across strata, worst-slab coverage in the
conformal literature \citep{romano2020,cauchois2021}, is
downward-biased by construction: the minimum of several noisy binomial
estimates falls below the true rate in expectation, by an amount that grows as
bins shrink. The bias is therefore largest exactly where materials datasets
sit. This is recognised in the methods literature, where implementations
mitigate it by splitting the evaluation sample and where \citet{braun2025}
replace the binned statistic with a classifier-based diagnostic. It is not
generally corrected for in applied materials UQ, including in the antecedent
study replicated here, which evaluated a 437-sample fatigue dataset on which a
method with exact conditional validity would have shown a worst-quartile
shortfall of 0.061 to 0.096 from sampling alone.

Two further measurement choices affect the same comparisons, and both have
been noted before in other settings. The effective quantile level rises as
per-group calibration sets shrink, since the finite-sample correction
$\lceil$(1-$\alpha$)(n+1)$\rceil$/n grows as 1/n; a group-conditional method calibrating within
small groups therefore operates above its nominal level and will appear
flatter across bins for reasons unrelated to conditional adaptivity
\citep{rafe2026}. And the resolution of the evaluation partition matters: a
group-conditional method has a guarantee defined on the partition it
calibrates on, so evaluating it there tests a promise made by construction,
while evaluating it on a refinement tests something the method never claimed.
This coarse-calibrate, fine-evaluate structure is the premise of clustered
conformal prediction \citep{ding2023}, which analyses the resulting
coverage gap in the many-class classification setting.

What this paper contributes is not these observations but their joint
application, at scale, in a domain where they are not currently applied and
where they change published conclusions. We replicate a single-dataset study
across five regression datasets spanning two orders of magnitude in sample
size, and address all three measurement issues explicitly: we report each
method's worst-bin shortfall against a null simulated from realized per-bin
counts and referenced to that method's realized marginal coverage; we evaluate
every method on the calibration partition, on a strict refinement of it, and on
a partition orthogonal to it; and we include a level-matched ablation
separating effective-level effects from conditional behaviour.

The corrections change the answer. Applied to the antecedent study's own
dataset they reverse its characterization of conformalized quantile
regression, and they split a method pair it had grouped together. The
replication's central positive finding, that group-conditional calibration
delivers what it promises on the partition it is given and measurably less on
that partition's refinement, is only visible once the null is subtracted,
because the uncorrected statistic assigns Mondrian a nonzero shortfall on
every dataset.

Section 2 describes the datasets, methods and evaluation partitions, and
defines the null-corrected statistic. Section 3 reports coverage, effective
level, null-corrected excess on all three partitions, and interval width.
Section 4 discusses what the corrections change, reconciles the divergences
with the antecedent study, positions the findings against the methods
literature above, and bounds their generality.

\subsection{Contributions}

\begin{enumerate}
\item A five-dataset, seven-method, 50-seed null-corrected replication of a single-dataset conditional coverage study in materials property prediction, which reverses two of its conclusions.
\item A level-matched ablation that separates a group-conditional method's effective quantile level from its conditional behaviour, on both the coverage statistic and the interval-width profile, and which shows its apparent width cost relative to split conformal prediction to be entirely attributable to the level.
\item A null for fixed-partition worst-bin coverage simulated from realized per-bin test counts rather than mitigated by sample splitting, referenced to each method's realized marginal coverage, with the diagnostic code released; and reporting recommendations for conditional coverage evaluation in small-sample applied UQ.
\end{enumerate}

\section{Methods}

\subsection{Datasets}

Five regression datasets were used, spanning materials property prediction and
one deliberately out-of-domain control. Table~\ref{tab:m1} lists sample sizes and target
properties.

\begin{table}[htbp]
\centering
\footnotesize
\caption{Datasets}
\label{tab:m1}
\begin{tabular}{lllp{0.36\textwidth}}
\toprule
Key & Target & n & Source \\
\midrule
\texttt{nims\_matnavi} & Rotating-bending fatigue strength & 437 & NIMS MatNavi \citep{furuya2019,nims2026}; preprocessed as in \citet{agrawal2014}; 25 composition/processing features \\
\texttt{uci\_concrete} & Compressive strength & 1030 & UCI id=165 \citep{yeh1998,yeh1998data} \\
\texttt{matbench\_expt\_gap} & Experimental band gap & 4604 & Matbench \citep{dunn2020}, data from \citet{zhuo2018}; Magpie-featurized \citep{ward2016,ward2018} \\
\texttt{california\_housing} & Median house value & 20640 & \citet{pace1997}, via scikit-learn \citep{pedregosa2011} \\
\texttt{uci\_superconductivity} & Critical temperature & 21263 & UCI id=464 \citep{hamidieh2018,hamidieh2018data}, 81 features \\
\bottomrule
\end{tabular}
\end{table}

\texttt{nims\_matnavi} is the dataset used in \citet{boruah2026} and serves both to validate the
port and as the first row of the replication. \texttt{uci\_superconductivity} was
selected as the primary replication target: same source institution as
\texttt{nims\_matnavi}, different property, two orders of magnitude larger.
\texttt{california\_housing} is not a materials dataset and is included as an
out-of-domain control.

\texttt{matbench\_expt\_gap} was featurized from composition using the Magpie preset of
\texttt{matminer}'s \texttt{ElementProperty} featurizer, with rows containing missing
features dropped. No further feature selection or transformation was
applied to any dataset.

\subsection{Split protocol and base model}

Each of 50 seeds defines an independent 60/20/20 train/calibration/test split,
generated by two successive calls to \texttt{train\_test\_split} (test\_size 0.4, then
0.5 on the holdout) with \texttt{random\_state} set to the seed. Features were
standardized using statistics fitted on the training split only. Realized
calibration and test sizes were 87/88 (\texttt{nims\_matnavi}), 206/206 (\texttt{uci\_concrete}),
921/921 (\texttt{matbench\_expt\_gap}), 4128/4128 (\texttt{california\_housing}), and 4253/4253
(\texttt{uci\_superconductivity}).

The antecedent study drew its test set first at 20\% and then sampled
calibration points from the remainder using a separate random generator.
Realized set sizes on \texttt{nims\_matnavi} are identical under both procedures
(262/87/88), but the two assign different points to each set, so the two
studies constitute independent 50-seed samples of the same estimand rather
than one being a rerun of the other. This accounts for the small differences
between the two papers' \texttt{nims\_matnavi} figures: the antecedent study's
top-quartile SCP coverage of 0.758 $\pm$ 0.109 has a standard error of 0.015
across its 50 seeds, and the 0.735 reported here (Table~\ref{tab:1}) falls within its
95\% confidence interval [0.728, 0.788], as do the corresponding marginal
coverage and mean interval width. Both studies bin on cut points taken from
out-of-fold training predictions, so the partition itself is not a source of
difference.

The base point model is a gradient boosting regressor \citep{friedman2001} with hyperparameters
frozen at the antecedent study's values across all datasets: 200 estimators, learning rate
0.05, max depth 4, minimum samples per leaf 1, subsample 1.0. Freezing these
rather than tuning per dataset keeps the interval-construction comparison from
being confounded by differences in point-model quality.

Out-of-fold training predictions were obtained by 5-fold cross-fitting with the
fold assignment fixed (\texttt{random\_state=0}, independent of the seed). These
predictions serve three purposes: the difficulty model for normalized
conformal prediction, the group boundaries for Mondrian calibration, and the
cut points for the evaluation partitions described below.

\subsection{Interval methods}

Seven methods were evaluated at a nominal miscoverage rate of $\alpha$ = 0.10.

\begin{itemize}
\item \textbf{\texttt{split\_conformal}} --- pooled absolute-residual quantile at the finite-sample-corrected level $\lceil$(1-$\alpha$)(n+1)$\rceil$/n, applied uniformly to all test points.
\item \textbf{\texttt{cross\_fitted\_normalized}} --- normalized conformal prediction \citep{papadopoulos2008,lei2018} with a difficulty model trained on out-of-fold absolute residuals, with a floor at 10\% of mean out-of-fold residual to prevent division by near-zero scale.
\item \textbf{\texttt{mondrian}} --- group-conditional split conformal \citep{vovk2003,bostrom2021}, with groups defined by quartiles of the out-of-fold training predictions and a separate corrected quantile computed within each group. Empty calibration groups fall back to the pooled quantile.
\item \textbf{\texttt{mondrian\_matched}} --- as above, but taking the interpolated empirical quantile at exactly 1-$\alpha$ with no finite-sample correction. This variant targets 1-$\alpha$ at every bin size and so is directly level-comparable to methods targeting 1-$\alpha$; it trades away the finite-sample coverage guarantee, and is included to isolate the conditional-coverage effect from the level offset rather than as a deployable method. Run on \texttt{nims\_matnavi} and \texttt{uci\_concrete} only, the two datasets where saturation is material.
\item \textbf{\texttt{cqr}} --- conformalized quantile regression \citep{romano2019}, with lower and upper quantile models at $\alpha$/2 and 1-$\alpha$/2 and a conformity correction from the calibration set.
\item \textbf{\texttt{raw\_quantile}} --- the same two quantile models \citep{koenker1978} without the conformal correction step. Included as an uncalibrated reference.
\item \textbf{\texttt{gpr}} --- Gaussian process regression \citep{rasmussen2006} with an RBF plus white-noise kernel, intervals from the predictive standard deviation at z = 1.645. Because GPR training is cubic in sample size, the training split was subsampled to 2000 points for \texttt{gpr} only on the three largest datasets; every other method saw the full split, and rows affected are flagged in the results.
\end{itemize}

A bootstrap ensemble method was evaluated in \citet{boruah2026} and is excluded here: it
fails the marginal validity precondition (0.650 marginal coverage on
\texttt{nims\_matnavi}) so its conditional behaviour is not interpretable, and its
runtime on the larger datasets is prohibitive.

\subsection{Evaluation partitions}

All conditional coverage is evaluated on partitions whose boundaries are
derived from training data only, and which are identical across methods. Both
constraints matter. Binning on each method's own predicted midpoint would give
every method a different partition, since \texttt{cqr}, \texttt{raw\_quantile} and \texttt{gpr} return
midpoints that differ from the base model's prediction, making columns
mutually incomparable. Deriving boundaries from test data would leak.

Three partitions are used:

\begin{itemize}
\item \textbf{Quartile} --- quartiles of the out-of-fold training predictions. This is the partition Mondrian calibrates on, so it is the on-partition, or as-specified, evaluation.
\item \textbf{Octile} --- octiles of the same out-of-fold training predictions. This is a refinement of the quartile partition on the identical variable, not an independent grouping: the 25th, 50th and 75th percentiles are by construction the 2nd, 4th and 6th octile cut points. Nesting was verified directly rather than assumed, by checking on 100 seeds (50 each on \texttt{nims\_matnavi} and \texttt{uci\_concrete}) that every quartile bin count equals the sum of its two constituent octile bin counts; there were no exceptions.
\item \textbf{Feature} --- quartiles of a single training feature, selected as the highest-importance feature having at least eight distinct values and strictly increasing quartile cut points. The eligibility rule is necessary because composition tables contain point masses: an element that is zero for most samples has identical quartile cut points, collapsing the partition to one or two non-empty bins. This partition conditions on a variable orthogonal to the one Mondrian calibrates on.
\end{itemize}

\subsection{The null-corrected min-gap statistic}

Conditional coverage failure is commonly reported as the minimum coverage
across evaluation bins, or its shortfall against nominal. This statistic is
biased: with finite test points per bin, the minimum of several noisy binomial
estimates falls below the true rate in expectation, so a method with exact
conditional validity will still appear to fail. The bias grows as bins shrink,
which makes it largest exactly where sample sizes are smallest and conclusions
are most fragile.

For each (dataset, method, partition, seed) we therefore report three
quantities:

\begin{itemize}
\item \textbf{\texttt{min\_gap}} = L - min\_b coverage\_b, the observed shortfall of the worst bin;
\item \textbf{null} = L - E[min\_b Binomial(n\_b, L)/n\_b], the expected value of the same statistic under exact conditional validity;
\item \textbf{excess} = \texttt{min\_gap} - null.
\end{itemize}

Each quantity is computed within a seed and then averaged across seeds. \texttt{min\_gap} is therefore the mean of the per-seed worst-bin shortfalls, not the shortfall of the worst column mean, and it does not in general equal the gap obtainable by subtracting the smallest quartile entry in Table~\ref{tab:1} from that row's marginal coverage. The two coincide only where the same bin is worst on nearly every seed.

The null is estimated by Monte Carlo (40,000 draws). Bin sizes matter because the map from bin size to expected minimum is nonlinear and convex: evaluating it at an equal split understates the null and so inflates every excess, by an amount that grows as the bins become more uneven. Relative unevenness scales as the inverse square root of the test-set size, so we used each seed's exact per-bin test counts on \texttt{nims\_matnavi} and \texttt{uci\_concrete}, where it is material, and an equal split on the three larger datasets, where simulating the realized counts moves the null by less than 0.0002 and no tabulated value. On \texttt{nims\_matnavi} the corresponding difference is 0.002, which is why the exact counts are used there. Those counts were recovered without refitting any interval method, by reproducing the split, scaler, base model and out-of-fold cut points for each seed --- these
determine the partitions and are independent of interval construction.

The reference level L is each row's own observed marginal coverage rather than
the nominal 1-$\alpha$. Conditioning on the method's realized marginal level
separates non-uniformity across bins from marginal miscalibration. Without it,
any method that misses marginally is charged for that miss on every
conditional test, regardless of how evenly its coverage is spread; in this
study that affected \texttt{mondrian\_matched}, whose marginal coverage is 0.865 and
0.882 by design, and \texttt{raw\_quantile}, which is uncalibrated. Results at the
nominal level are reported alongside for comparison in Table~\ref{tab:4b}.

Because the excess distribution is right-skewed, we report both the mean excess across seeds and the fraction of seeds on which \texttt{min\_gap} exceeded the 95th percentile of that seed's null.

Three quantities therefore describe each cell and should not be conflated. The mean excess with its interval (Table~\ref{tab:4c}) says how large the departure from the null is. The per-seed exceedance rate says how often a single split would have flagged it, against 0.05 under exact conditional validity. The majority-of-seeds criterion is a deliberately conservative summary of the second, used for the dataset-level counts in Section 4. The first two are the substantive quantities; the third is a reporting convention and nothing in Section 4 turns on where its line is drawn. The intervals in Table~\ref{tab:4c} are computed across seeds as though the splits were independent. They overlap substantially, so those intervals are narrower than the design warrants, in the same direction and for the same reason as the exceedance counts discussed below.

Under exact conditional validity that rate is 0.05 by construction, so the
majority-of-seeds criterion applied below is an order of magnitude more
conservative than a per-seed test. We use it because the 50 splits are drawn
from one fixed dataset and overlap substantially: exceedance counts across
seeds are positively correlated, and a binomial test treating them as
independent would be anticonservative. Rates that fall below the criterion
but sit materially above 0.05 are reported and discussed rather than treated
as null, and the tables below state the 0.05 reference so that individual
cells can be read against it directly.

\subsection{Effective quantile level}

Mondrian's per-bin quantile uses the corrected index $\lceil$(1-$\alpha$)(n\_b+1)$\rceil$/n\_b, which
rises as bin size falls and saturates at 1.0 once the required order statistic
exceeds the bin count. At saturation the interval is the largest calibration
residual in the bin: still conservative, but no longer described by the nominal
level. Per-bin sizes, realized levels and saturation flags were logged for
every bin and seed. Three summaries are reported and should not be conflated: the
mean level across bins, which describes the level Mondrian operates at on
average; the mean of the per-seed maximum bin level, which describes its worst
bin; and the pooled level, the single corrected level
$\lceil$(1-$\alpha$)(n\_cal+1)$\rceil$/n\_cal that split conformal prediction applies to the whole
calibration set, included as the reference against which the per-bin inflation
should be read.

\subsection{Reproducibility}

All results derive from a single canonical run. Seed counts and row counts were
checked for every dataset before analysis, since the result-writing path
appends rather than overwrites and a partially repeated run can produce a file
that passes a naive seed-count check. Realized row counts were 300 (six
methods $\times$ 50 seeds) on the three datasets without the matched variant and 350
(seven methods $\times$ 50 seeds) on \texttt{nims\_matnavi} and \texttt{uci\_concrete}, with no duplicated
(method, seed) pairs and no uneven seed counts.

\section{Results}

\subsection{Coverage on the calibration partition}

Table~\ref{tab:1} reports empirical coverage within each prediction quartile, and
marginal coverage, for all seven methods on all five datasets, against a
nominal target of 0.90. The two methods the rest of this section returns to
most often are split conformal prediction (SCP), the constant-width reference,
and group-conditional (Mondrian) conformal prediction.

\begin{table}[htbp]
\centering
\footnotesize
\caption{Coverage by prediction quartile and marginal coverage, all methods, 50-seed means (target 0.90)}
\label{tab:1}
\begin{tabular}{llrrrrr}
\toprule
Dataset & Method & Q1 & Q2 & Q3 & Q4 & Marginal \\
\midrule
\texttt{nims\_matnavi} & \texttt{cqr} & 0.928 & 0.984 & 0.970 & 0.797 & 0.917 \\
\texttt{nims\_matnavi} & \texttt{cross\_fitted\_normalized} & 0.927 & 0.946 & 0.907 & 0.858 & 0.909 \\
\texttt{nims\_matnavi} & \texttt{gpr} & 0.982 & 0.937 & 0.885 & 0.732 & 0.882 \\
\texttt{nims\_matnavi} & \texttt{mondrian} & 0.926 & 0.915 & 0.922 & 0.925 & 0.921 \\
\texttt{nims\_matnavi} & \texttt{mondrian\_matched} & 0.867 & 0.865 & 0.857 & 0.864 & 0.865 \\
\texttt{nims\_matnavi} & \texttt{raw\_quantile} & 0.667 & 0.888 & 0.866 & 0.672 & 0.770 \\
\texttt{nims\_matnavi} & \texttt{split\_conformal} & 0.980 & 0.969 & 0.972 & 0.735 & 0.912 \\
\texttt{uci\_concrete} & \texttt{cqr} & 0.891 & 0.949 & 0.938 & 0.823 & 0.900 \\
\texttt{uci\_concrete} & \texttt{cross\_fitted\_normalized} & 0.924 & 0.898 & 0.888 & 0.892 & 0.900 \\
\texttt{uci\_concrete} & \texttt{gpr} & 0.955 & 0.897 & 0.863 & 0.821 & 0.883 \\
\texttt{uci\_concrete} & \texttt{mondrian} & 0.913 & 0.915 & 0.899 & 0.903 & 0.907 \\
\texttt{uci\_concrete} & \texttt{mondrian\_matched} & 0.888 & 0.893 & 0.876 & 0.874 & 0.882 \\
\texttt{uci\_concrete} & \texttt{raw\_quantile} & 0.696 & 0.884 & 0.876 & 0.668 & 0.782 \\
\texttt{uci\_concrete} & \texttt{split\_conformal} & 0.964 & 0.918 & 0.900 & 0.827 & 0.902 \\
\texttt{matbench\_expt\_gap} & \texttt{cqr} & 0.980 & 0.967 & 0.938 & 0.842 & 0.932 \\
\texttt{matbench\_expt\_gap} & \texttt{cross\_fitted\_normalized} & 0.974 & 0.911 & 0.836 & 0.876 & 0.900 \\
\texttt{matbench\_expt\_gap} & \texttt{gpr} & 0.995 & 0.964 & 0.878 & 0.804 & 0.911 \\
\texttt{matbench\_expt\_gap} & \texttt{mondrian} & 0.896 & 0.906 & 0.905 & 0.904 & 0.903 \\
\texttt{matbench\_expt\_gap} & \texttt{raw\_quantile} & 0.980 & 0.967 & 0.938 & 0.842 & 0.932 \\
\texttt{matbench\_expt\_gap} & \texttt{split\_conformal} & 0.993 & 0.970 & 0.866 & 0.777 & 0.902 \\
\texttt{california\_housing} & \texttt{cqr} & 0.902 & 0.898 & 0.905 & 0.897 & 0.901 \\
\texttt{california\_housing} & \texttt{cross\_fitted\_normalized} & 0.950 & 0.912 & 0.881 & 0.858 & 0.900 \\
\texttt{california\_housing} & \texttt{gpr} & 0.982 & 0.953 & 0.903 & 0.814 & 0.913 \\
\texttt{california\_housing} & \texttt{mondrian} & 0.902 & 0.899 & 0.900 & 0.899 & 0.900 \\
\texttt{california\_housing} & \texttt{raw\_quantile} & 0.878 & 0.886 & 0.899 & 0.811 & 0.868 \\
\texttt{california\_housing} & \texttt{split\_conformal} & 0.981 & 0.946 & 0.879 & 0.792 & 0.900 \\
\texttt{uci\_superconductivity} & \texttt{cqr} & 0.915 & 0.896 & 0.887 & 0.901 & 0.899 \\
\texttt{uci\_superconductivity} & \texttt{cross\_fitted\_normalized} & 0.957 & 0.898 & 0.835 & 0.909 & 0.900 \\
\texttt{uci\_superconductivity} & \texttt{gpr} & 0.999 & 0.989 & 0.792 & 0.843 & 0.905 \\
\texttt{uci\_superconductivity} & \texttt{mondrian} & 0.902 & 0.899 & 0.900 & 0.904 & 0.901 \\
\texttt{uci\_superconductivity} & \texttt{raw\_quantile} & 0.885 & 0.883 & 0.882 & 0.895 & 0.886 \\
\texttt{uci\_superconductivity} & \texttt{split\_conformal} & 0.999 & 0.986 & 0.778 & 0.841 & 0.901 \\
\bottomrule
\end{tabular}
\end{table}

\texttt{mondrian\_matched} was run on \texttt{nims\_matnavi} and \texttt{uci\_concrete} only (Methods).
Marginal coverage was within 0.02 of nominal for every method on every dataset
except \texttt{raw\_quantile}, which is uncalibrated by construction, and
\texttt{mondrian\_matched}, whose level is reduced by design.

SCP undercovered in Q4 on all five datasets, by 0.06 to 0.17. Mondrian's Q4
coverage sat within 0.03 of nominal everywhere. SCP's worst quartile is Q4 on
four datasets and Q3 on \texttt{uci\_superconductivity} (0.778 against 0.841 in Q4);
\texttt{cross\_fitted\_normalized}'s worst is likewise Q3 on \texttt{uci\_superconductivity}
(0.835) and on \texttt{matbench\_expt\_gap} (0.836) rather than Q4. On
\texttt{matbench\_expt\_gap}, \texttt{cqr} and \texttt{raw\_quantile} are identical in every cell, for the
reason given under "Dataset-specific observations".

\subsection{Effective Mondrian level}

Mondrian's per-bin quantile uses the finite-sample-corrected index
$\lceil$(1-$\alpha$)(n\_b+1)$\rceil$/n\_b, which rises as bin size n\_b falls and saturates at 1.0 once
the corrected count reaches n\_b. Table~\ref{tab:2} reports, for each dataset, the
calibration set size, the smallest observed per-bin count across 50 seeds, the
mean effective level averaged across the four bins, the mean of the per-seed
maximum bin level, and the pooled level that split conformal prediction
operates at on the same calibration set.

\begin{table}[htbp]
\centering
\footnotesize
\caption{Effective Mondrian quantile level}
\label{tab:2}
\begin{tabular}{llrrrr}
\toprule
Dataset & n\_cal & min bin n & mean level (across bins) & mean level (worst bin) & pooled level \\
\midrule
\texttt{nims\_matnavi} & 87 & 10 & 0.965 & 0.993 & 0.9195 \\
\texttt{uci\_concrete} & 206 & 34 & 0.926 & 0.934 & 0.9078 \\
\texttt{matbench\_expt\_gap} & 921 & 192 & 0.906 & 0.907 & 0.9012 \\
\texttt{california\_housing} & 4128 & 933 & 0.901 & 0.902 & 0.9004 \\
\texttt{uci\_superconductivity} & 4253 & 966 & 0.901 & 0.902 & 0.9003 \\
\bottomrule
\end{tabular}
\end{table}

On \texttt{nims\_matnavi}, 41 of 50 seeds produced at least one saturated bin (level =
1.0), and 52 of the 200 bin-seed combinations were saturated (26.0\%). No
saturation occurred on the other four datasets.

A level-matched variant of Mondrian, using the interpolated empirical quantile
at exactly 1-$\alpha$ with no finite-sample correction, was run on the two datasets
where saturation was material (\texttt{nims\_matnavi}, \texttt{uci\_concrete}). Marginal coverage
for this variant was 0.865 on \texttt{nims\_matnavi} and 0.882 on \texttt{uci\_concrete}, close to the order-statistic prediction (0.9(n-1)+1)/(n+1) evaluated at the mean bin sizes observed for each dataset, which gives 0.865 at 21.75 and 0.885 at 51.5. That prediction is exact on \texttt{nims\_matnavi} and 0.003 high on \texttt{uci\_concrete}, as expected: coverage is concave in bin size, so averaging it over the realized spread of bin sizes falls below its value at the mean. The quartile coverage profile
for the level-matched variant remained flat across bins (\texttt{nims\_matnavi}: 0.867,
0.865, 0.857, 0.864), indicating that Mondrian's within-partition uniformity is
not an artifact of its elevated effective level.

\subsection{A null-corrected statistic for min-over-bins coverage}

Reporting the minimum coverage across k evaluation bins understates the target
level even when a method is exactly conditionally valid: with finite test
samples per bin, the minimum of k noisy binomial estimates falls below the
nominal rate in expectation. This section reports, for each (dataset, method,
partition), the observed shortfall of the worst bin against that method's realized marginal coverage (\texttt{min\_gap}), the expected value of that same statistic under
exact conditional validity at the method's own observed marginal coverage
(null), and their difference (excess = \texttt{min\_gap} - null). The null was estimated
by Monte Carlo simulation (40,000 draws) using the exact per-bin test counts
recorded for each seed. Conditioning the null on each row's observed marginal
coverage, rather than on the nominal 1-$\alpha$, separates non-uniformity across bins
from marginal miscalibration; a method that undercovers marginally will
otherwise appear to fail every conditional test regardless of how evenly its
coverage is distributed across bins. Both quantities are reported per seed and
averaged; the fraction of seeds on which \texttt{min\_gap} exceeded the null's 95th
percentile is reported alongside the mean, since the excess statistic is
right-skewed.

Three partitions were evaluated for every method: the quartile partition
Mondrian calibrates on, an octile refinement of the same variable (eight bins
from the same out-of-fold training predictions used for the quartile cuts),
and a partition on the training feature with the highest importance among
features carrying at least eight distinct values and strictly increasing
quartile cut points. The quartile partition is a strict coarsening of the
octile one: bin b of the quartile partition is the union of octile bins 2b
and 2b+1, verified directly against all 100 seeds (50 each on
\texttt{nims\_matnavi} and \texttt{uci\_concrete}) with zero exceptions.

\subsubsection{On-partition (quartile)}

\begin{table}[htbp]
\centering
\footnotesize
\caption{The null itself: expected worst-quartile shortfall under exact conditional validity, at each method's own realized marginal coverage}
\label{tab:3}
\begin{tabular}{llrrrr}
\toprule
Method & california & matbench & matnavi & concrete & supercon \\
\midrule
\texttt{cqr} & 0.010 & 0.017 & 0.063 & 0.044 & 0.010 \\
\texttt{cross\_fitted\_normalized} & 0.010 & 0.021 & 0.066 & 0.044 & 0.010 \\
\texttt{gpr} & 0.009 & 0.020 & 0.074 & 0.047 & 0.009 \\
\texttt{mondrian} & 0.010 & 0.020 & 0.061 & 0.043 & 0.009 \\
\texttt{mondrian\_matched} & --- & --- & 0.078 & 0.047 & --- \\
\texttt{raw\_quantile} & 0.011 & 0.017 & 0.096 & 0.061 & 0.010 \\
\texttt{split\_conformal} & 0.010 & 0.020 & 0.065 & 0.044 & 0.009 \\
\bottomrule
\end{tabular}
\end{table}

The null spans an order of magnitude across datasets, from 0.009--0.011 on the
two largest to 0.061--0.096 on \texttt{nims\_matnavi}, tracking test-set size. Within a
dataset it also varies by method, because it is evaluated at each method's own
marginal coverage: \texttt{raw\_quantile} and \texttt{mondrian\_matched}, which operate below
0.90, have larger nulls than methods calibrated at the nominal level, since
binomial dispersion increases as the reference level falls away from 1.

The correction is therefore not uniform in importance. On \texttt{california\_housing},
\texttt{split\_conformal}'s raw worst-quartile shortfall of 0.108 is 91\% real signal;
subtracting the null changes the conclusion little. On \texttt{nims\_matnavi}, \texttt{mondrian}'s
raw shortfall of 0.094 is 65\% null, \texttt{cqr}'s 0.124 is 51\% null, and
\texttt{split\_conformal}'s 0.177 is 37\% null. The uncorrected statistic is close to
usable on large datasets and actively misleading on small ones.

\begin{table}[htbp]
\centering
\footnotesize
\caption{Null-corrected excess, quartile partition ($\alpha$ = 0.10)}
\label{tab:4}
\begin{tabular}{llrrrr}
\toprule
Method & california & matbench & matnavi & concrete & supercon \\
\midrule
\texttt{cqr} & 0.003 & 0.073 & 0.061 & 0.042 & 0.007 \\
\texttt{cross\_fitted\_normalized} & 0.033 & 0.045 & 0.019 & 0.008 & 0.055 \\
\texttt{gpr} & 0.090 & 0.088 & 0.078 & 0.030 & 0.104 \\
\texttt{mondrian} & 0.003 & 0.010 & 0.033 & 0.021 & 0.006 \\
\texttt{mondrian\_matched} & --- & --- & 0.039 & 0.020 & --- \\
\texttt{raw\_quantile} & 0.048 & 0.073 & 0.065 & 0.075 & 0.003 \\
\texttt{split\_conformal} & 0.098 & 0.105 & 0.112 & 0.036 & 0.113 \\
\bottomrule
\end{tabular}
\end{table}

GPR was trained on a 2000-point subsample of the training split on
\texttt{california\_housing}, \texttt{matbench\_expt\_gap} and \texttt{uci\_superconductivity} (Methods).
Those are the three datasets on which it exceeds the null on a majority of
seeds below; its rows there reflect a model fitted on a fraction of the
training data available to every other method, and are flagged again wherever
they are compared.

\begin{table}[htbp]
\centering
\footnotesize
\caption{Null-corrected quartile excess with the half-width of its 95\% interval (1.96 standard errors across seeds); the interval is the excess plus and minus the printed half-width. Intervals treat the 50 splits as independent and so are narrower than the overlap between them warrants; see Methods}
\label{tab:4c}
\resizebox{\textwidth}{!}{%
\begin{tabular}{llrrrr}
\toprule
Method & california & matbench & matnavi & concrete & supercon \\
\midrule
\texttt{cqr} & 0.003 $\pm$0.002 & 0.073 $\pm$0.006 & 0.061 $\pm$0.013 & 0.042 $\pm$0.011 & 0.007 $\pm$0.002 \\
\texttt{cross\_fitted\_normalized} & 0.033 $\pm$0.002 & 0.045 $\pm$0.005 & 0.019 $\pm$0.012 & 0.008 $\pm$0.008 & 0.055 $\pm$0.003 \\
\texttt{gpr} & 0.090 $\pm$0.003 & 0.088 $\pm$0.006 & 0.078 $\pm$0.017 & 0.030 $\pm$0.009 & 0.104 $\pm$0.004 \\
\texttt{mondrian} & 0.003 $\pm$0.002 & 0.010 $\pm$0.005 & 0.033 $\pm$0.013 & 0.021 $\pm$0.009 & 0.006 $\pm$0.002 \\
\texttt{mondrian\_matched} & --- & --- & 0.039 $\pm$0.018 & 0.020 $\pm$0.010 & --- \\
\texttt{raw\_quantile} & 0.048 $\pm$0.004 & 0.073 $\pm$0.006 & 0.065 $\pm$0.017 & 0.075 $\pm$0.011 & 0.003 $\pm$0.002 \\
\texttt{split\_conformal} & 0.098 $\pm$0.003 & 0.105 $\pm$0.007 & 0.112 $\pm$0.021 & 0.036 $\pm$0.011 & 0.113 $\pm$0.004 \\
\bottomrule
\end{tabular}
}
\end{table}

Every interval excludes zero, including \texttt{mondrian}'s, so no method in this study is exactly conditionally valid at quartile resolution. What separates them is magnitude: on the three largest datasets \texttt{mondrian}'s excess is between a thirtieth and a tenth of \texttt{split\_conformal}'s, and its interval does not approach the latter's on any dataset.

\begin{table}[htbp]
\centering
\footnotesize
\caption{Null-corrected excess on the quartile partition with the null referenced to the nominal 0.90 instead of each method's realized marginal coverage, shown against Table~\ref{tab:4} and with the realized marginal level. Differences are computed at full precision and rounded independently of the two columns beside them, so they need not equal the difference of the rounded values}
\label{tab:4b}
\begin{tabular}{llrrrr}
\toprule
Dataset & Method & excess (nominal) & excess (marginal) & diff & marginal level \\
\midrule
california & \texttt{cqr} & 0.001 & 0.003 & -0.001 & 0.901 \\
california & \texttt{cross\_fitted\_normalized} & 0.033 & 0.033 & -0.000 & 0.900 \\
california & \texttt{gpr} & 0.076 & 0.090 & -0.014 & 0.913 \\
california & \texttt{mondrian} & 0.003 & 0.003 & -0.000 & 0.900 \\
california & \texttt{raw\_quantile} & 0.080 & 0.048 & +0.032 & 0.868 \\
california & \texttt{split\_conformal} & 0.098 & 0.098 & +0.000 & 0.900 \\
matbench & \texttt{cqr} & 0.036 & 0.073 & -0.037 & 0.932 \\
matbench & \texttt{cross\_fitted\_normalized} & 0.043 & 0.045 & -0.002 & 0.900 \\
matbench & \texttt{gpr} & 0.074 & 0.088 & -0.013 & 0.911 \\
matbench & \texttt{mondrian} & 0.005 & 0.010 & -0.005 & 0.903 \\
matbench & \texttt{raw\_quantile} & 0.036 & 0.073 & -0.037 & 0.932 \\
matbench & \texttt{split\_conformal} & 0.101 & 0.105 & -0.004 & 0.902 \\
matnavi & \texttt{cqr} & 0.015 & 0.061 & -0.046 & 0.917 \\
matnavi & \texttt{cross\_fitted\_normalized} & -0.016 & 0.019 & -0.036 & 0.909 \\
matnavi & \texttt{gpr} & 0.078 & 0.078 & -0.001 & 0.882 \\
matnavi & \texttt{mondrian} & -0.020 & 0.033 & -0.052 & 0.921 \\
matnavi & \texttt{mondrian\_matched} & 0.060 & 0.039 & +0.021 & 0.865 \\
matnavi & \texttt{raw\_quantile} & 0.198 & 0.065 & +0.133 & 0.770 \\
matnavi & \texttt{split\_conformal} & 0.073 & 0.112 & -0.040 & 0.912 \\
concrete & \texttt{cqr} & 0.033 & 0.042 & -0.009 & 0.900 \\
concrete & \texttt{cross\_fitted\_normalized} & -0.001 & 0.008 & -0.009 & 0.900 \\
concrete & \texttt{gpr} & 0.040 & 0.030 & +0.011 & 0.883 \\
concrete & \texttt{mondrian} & 0.003 & 0.021 & -0.017 & 0.907 \\
concrete & \texttt{mondrian\_matched} & 0.032 & 0.020 & +0.012 & 0.882 \\
concrete & \texttt{raw\_quantile} & 0.200 & 0.075 & +0.125 & 0.782 \\
concrete & \texttt{split\_conformal} & 0.025 & 0.036 & -0.011 & 0.902 \\
supercon & \texttt{cqr} & 0.007 & 0.007 & +0.000 & 0.899 \\
supercon & \texttt{cross\_fitted\_normalized} & 0.055 & 0.055 & -0.000 & 0.900 \\
supercon & \texttt{gpr} & 0.098 & 0.104 & -0.006 & 0.905 \\
supercon & \texttt{mondrian} & 0.005 & 0.006 & -0.002 & 0.901 \\
supercon & \texttt{raw\_quantile} & 0.017 & 0.003 & +0.014 & 0.886 \\
supercon & \texttt{split\_conformal} & 0.112 & 0.113 & -0.001 & 0.901 \\
\bottomrule
\end{tabular}
\end{table}

The two references agree wherever a method sits at nominal. They diverge in
proportion to how far its realized marginal coverage departs from it: at or below 0.002 on the two largest datasets for methods whose realized marginal coverage sits within 0.002 of nominal, rising to 0.004 on \texttt{matbench\_expt\_gap} and 0.011 on \texttt{uci\_concrete} for equally well-calibrated methods, since the null itself is larger there and so more sensitive to the reference level; and growing in both directions as realized marginal coverage departs from nominal. \texttt{raw\_quantile}, marginally
invalid at 0.770 and 0.782, is charged 0.133 and 0.125 more at the nominal
reference than at its own, since the nominal reference attributes its marginal
miss to the conditional statistic. Over-covering methods move the other way:
at the nominal reference \texttt{mondrian} returns a negative excess on \texttt{nims\_matnavi}
(-0.020) and \texttt{cross\_fitted\_normalized} on both smallest datasets (-0.016,
-0.001), because a method covering above 0.90 marginally clears a null built
at 0.90 in every bin. Neither direction is informative about uniformity across
bins, which is why the marginal reference is used throughout Tables~\ref{tab:3} to~\ref{tab:9}. The off-partition rows follow the same pattern with larger magnitudes, peaking at 0.179 for \texttt{raw\_quantile} on the \texttt{nims\_matnavi} octile partition; this figure is taken from the harness's stored nominal null at 20,000 draws rather than recomputed at 40,000 like the marginal-referenced values, a difference well below its leading digit.

\begin{table}[htbp]
\centering
\footnotesize
\caption{Fraction of seeds with \texttt{min\_gap} above the null's 95th percentile, quartile partition. Under exact conditional validity this rate is 0.05}
\label{tab:5}
\begin{tabular}{llrrrr}
\toprule
Method & california & matbench & matnavi & concrete & supercon \\
\midrule
\texttt{cqr} & 0.10 & 0.98 & 0.24 & 0.24 & 0.30 \\
\texttt{cross\_fitted\_normalized} & 1.00 & 0.82 & 0.04 & 0.08 & 1.00 \\
\texttt{gpr} & 1.00 & 1.00 & 0.38 & 0.16 & 1.00 \\
\texttt{mondrian} & 0.10 & 0.20 & 0.12 & 0.14 & 0.24 \\
\texttt{mondrian\_matched} & --- & --- & 0.18 & 0.20 & --- \\
\texttt{raw\_quantile} & 0.98 & 0.98 & 0.20 & 0.48 & 0.08 \\
\texttt{split\_conformal} & 1.00 & 1.00 & 0.60 & 0.24 & 1.00 \\
\bottomrule
\end{tabular}
\end{table}

Treating a method as failing on a dataset when its \texttt{min\_gap} exceeded the null's
95th percentile on more than half of the 50 seeds, \texttt{split\_conformal} failed on
four of five datasets, and \texttt{gpr} and \texttt{cross\_fitted\_normalized} on three each.
\texttt{mondrian} and \texttt{mondrian\_matched} failed on none (peak significance rate 0.24, on
superconductivity). That peak is not zero: it sits roughly five times the 0.05
rate expected under exact conditional validity, and is read here as a small
on-partition deficit below the study's criterion rather than as an absence of
one. \texttt{cqr} failed on one dataset (\texttt{matbench\_expt\_gap}, 0.98); the CQR/\texttt{raw\_quantile}
identity on that dataset is reported below. \texttt{raw\_quantile} failed on two.

\subsubsection{Off-partition (octile and feature)}

\begin{table}[htbp]
\centering
\footnotesize
\caption{Null-corrected excess, octile partition}
\label{tab:6}
\begin{tabular}{llrrrr}
\toprule
Method & california & matbench & matnavi & concrete & supercon \\
\midrule
\texttt{cqr} & 0.006 & 0.110 & 0.055 & 0.043 & 0.017 \\
\texttt{cross\_fitted\_normalized} & 0.035 & 0.049 & 0.018 & 0.010 & 0.082 \\
\texttt{gpr} & 0.101 & 0.139 & 0.079 & 0.038 & 0.209 \\
\texttt{mondrian} & 0.021 & 0.074 & 0.029 & 0.027 & 0.085 \\
\texttt{mondrian\_matched} & --- & --- & 0.045 & 0.031 & --- \\
\texttt{raw\_quantile} & 0.115 & 0.110 & 0.098 & 0.092 & 0.012 \\
\texttt{split\_conformal} & 0.097 & 0.136 & 0.164 & 0.044 & 0.230 \\
\bottomrule
\end{tabular}
\end{table}

\begin{table}[htbp]
\centering
\footnotesize
\caption{Null-corrected excess on the octile partition with the half-width of its 95\% interval (1.96 standard errors across seeds); the interval is the excess plus and minus the printed half-width. Intervals treat the 50 splits as independent and so are narrower than the overlap between them warrants; see Methods}
\label{tab:6c}
\resizebox{\textwidth}{!}{%
\begin{tabular}{llrrrr}
\toprule
Method & california & matbench & matnavi & concrete & supercon \\
\midrule
\texttt{cqr} & 0.006 $\pm$0.002 & 0.110 $\pm$0.009 & 0.055 $\pm$0.019 & 0.043 $\pm$0.014 & 0.017 $\pm$0.003 \\
\texttt{cross\_fitted\_normalized} & 0.035 $\pm$0.003 & 0.049 $\pm$0.007 & 0.018 $\pm$0.021 & 0.010 $\pm$0.011 & 0.082 $\pm$0.005 \\
\texttt{gpr} & 0.101 $\pm$0.005 & 0.139 $\pm$0.010 & 0.079 $\pm$0.024 & 0.038 $\pm$0.015 & 0.209 $\pm$0.007 \\
\texttt{mondrian} & 0.021 $\pm$0.003 & 0.074 $\pm$0.009 & 0.029 $\pm$0.022 & 0.027 $\pm$0.013 & 0.085 $\pm$0.005 \\
\texttt{mondrian\_matched} & --- & --- & 0.045 $\pm$0.028 & 0.031 $\pm$0.012 & --- \\
\texttt{raw\_quantile} & 0.115 $\pm$0.008 & 0.110 $\pm$0.009 & 0.098 $\pm$0.028 & 0.092 $\pm$0.018 & 0.012 $\pm$0.003 \\
\texttt{split\_conformal} & 0.097 $\pm$0.003 & 0.136 $\pm$0.010 & 0.164 $\pm$0.030 & 0.044 $\pm$0.013 & 0.230 $\pm$0.007 \\
\bottomrule
\end{tabular}
}
\end{table}

\begin{table}[htbp]
\centering
\footnotesize
\caption{Fraction of seeds above the null's 95th percentile, octile partition. Under exact conditional validity this rate is 0.05}
\label{tab:7}
\begin{tabular}{llrrrr}
\toprule
Method & california & matbench & matnavi & concrete & supercon \\
\midrule
\texttt{cqr} & 0.18 & 1.00 & 0.10 & 0.18 & 0.50 \\
\texttt{cross\_fitted\_normalized} & 1.00 & 0.72 & 0.06 & 0.02 & 1.00 \\
\texttt{gpr} & 1.00 & 1.00 & 0.14 & 0.16 & 1.00 \\
\texttt{mondrian} & 0.66 & 0.96 & 0.10 & 0.10 & 1.00 \\
\texttt{mondrian\_matched} & --- & --- & 0.14 & 0.12 & --- \\
\texttt{raw\_quantile} & 0.98 & 1.00 & 0.24 & 0.42 & 0.34 \\
\texttt{split\_conformal} & 1.00 & 1.00 & 0.52 & 0.18 & 1.00 \\
\bottomrule
\end{tabular}
\end{table}

\texttt{mondrian} exceeded the null on the octile partition on three of five datasets
(california, \texttt{matbench\_expt\_gap}, superconductivity; significance rate $\geq$ 0.66),
against zero on the coarser quartile partition. On \texttt{nims\_matnavi} and
\texttt{uci\_concrete}, the two smallest datasets, \texttt{mondrian} did not exceed the null on
either partition (octile significance rate 0.10). The smallest observed
octile bin on \texttt{nims\_matnavi} held as few as one test point in a given seed
(range 1--9 across seeds); the smallest observed octile bin on \texttt{uci\_concrete}
held 11--23 points.

Feature-partition results (Table~\ref{tab:8}, Table~\ref{tab:9}) followed a similar pattern to
the octile partition, with lower excess values overall. The partitioning
feature selected for each dataset was: \texttt{nims\_matnavi} → Cr; \texttt{uci\_concrete} → Age;
\texttt{california\_housing} → MedInc; \texttt{matbench\_expt\_gap} → MagpieData mean NpValence;
\texttt{uci\_superconductivity} → range\_ThermalConductivity.

\begin{table}[htbp]
\centering
\footnotesize
\caption{Null-corrected excess, feature partition}
\label{tab:8}
\begin{tabular}{llrrrr}
\toprule
Method & california & matbench & matnavi & concrete & supercon \\
\midrule
\texttt{cqr} & 0.003 & 0.013 & 0.023 & 0.006 & -0.002 \\
\texttt{cross\_fitted\_normalized} & 0.008 & 0.015 & 0.016 & 0.001 & -0.008 \\
\texttt{gpr} & 0.018 & 0.053 & 0.015 & 0.029 & 0.064 \\
\texttt{mondrian} & 0.015 & 0.014 & 0.036 & 0.015 & 0.059 \\
\texttt{mondrian\_matched} & --- & --- & 0.044 & 0.016 & --- \\
\texttt{raw\_quantile} & 0.041 & 0.013 & 0.033 & 0.019 & 0.003 \\
\texttt{split\_conformal} & 0.026 & 0.050 & 0.038 & -0.004 & 0.055 \\
\bottomrule
\end{tabular}
\end{table}

\begin{table}[htbp]
\centering
\footnotesize
\caption{Null-corrected excess on the feature partition with the half-width of its 95\% interval (1.96 standard errors across seeds); the interval is the excess plus and minus the printed half-width. Intervals treat the 50 splits as independent and so are narrower than the overlap between them warrants; see Methods}
\label{tab:8c}
\resizebox{\textwidth}{!}{%
\begin{tabular}{llrrrr}
\toprule
Method & california & matbench & matnavi & concrete & supercon \\
\midrule
\texttt{cqr} & 0.003 $\pm$0.002 & 0.013 $\pm$0.003 & 0.023 $\pm$0.014 & 0.006 $\pm$0.011 & -0.002 $\pm$0.009 \\
\texttt{cross\_fitted\_normalized} & 0.008 $\pm$0.002 & 0.015 $\pm$0.005 & 0.016 $\pm$0.011 & 0.001 $\pm$0.010 & -0.008 $\pm$0.003 \\
\texttt{gpr} & 0.018 $\pm$0.002 & 0.053 $\pm$0.004 & 0.015 $\pm$0.009 & 0.029 $\pm$0.013 & 0.064 $\pm$0.004 \\
\texttt{mondrian} & 0.015 $\pm$0.002 & 0.014 $\pm$0.003 & 0.036 $\pm$0.016 & 0.015 $\pm$0.010 & 0.059 $\pm$0.022 \\
\texttt{mondrian\_matched} & --- & --- & 0.044 $\pm$0.016 & 0.016 $\pm$0.013 & --- \\
\texttt{raw\_quantile} & 0.041 $\pm$0.004 & 0.013 $\pm$0.003 & 0.033 $\pm$0.018 & 0.019 $\pm$0.016 & 0.003 $\pm$0.010 \\
\texttt{split\_conformal} & 0.026 $\pm$0.002 & 0.050 $\pm$0.004 & 0.038 $\pm$0.013 & -0.004 $\pm$0.006 & 0.055 $\pm$0.003 \\
\bottomrule
\end{tabular}
}
\end{table}

\begin{table}[htbp]
\centering
\footnotesize
\caption{Fraction of seeds above the null's 95th percentile, feature partition. Under exact conditional validity this rate is 0.05}
\label{tab:9}
\begin{tabular}{llrrrr}
\toprule
Method & california & matbench & matnavi & concrete & supercon \\
\midrule
\texttt{cqr} & 0.08 & 0.20 & 0.08 & 0.04 & 0.08 \\
\texttt{cross\_fitted\_normalized} & 0.32 & 0.22 & 0.06 & 0.02 & 0.08 \\
\texttt{gpr} & 0.90 & 0.98 & 0.00 & 0.18 & 0.34 \\
\texttt{mondrian} & 0.62 & 0.10 & 0.14 & 0.06 & 0.30 \\
\texttt{mondrian\_matched} & --- & --- & 0.10 & 0.06 & --- \\
\texttt{raw\_quantile} & 0.96 & 0.20 & 0.10 & 0.06 & 0.14 \\
\texttt{split\_conformal} & 1.00 & 0.96 & 0.08 & 0.00 & 0.30 \\
\bottomrule
\end{tabular}
\end{table}

\texttt{cqr} and \texttt{cross\_fitted\_normalized} had the lowest excess of any method on the feature partition on three of the five datasets, and came within 0.005 of the lowest on the other two (excess $\leq$ 0.023, significance rate
$\leq$ 0.32 throughout), including two negative excess values (\texttt{cqr}, superconductivity,
-0.002; \texttt{cross\_fitted\_normalized}, superconductivity, -0.008). Negative excess means the observed worst-bin shortfall fell below its expectation under exact conditional validity. For \texttt{cqr} on \texttt{uci\_superconductivity} and \texttt{split\_conformal} on \texttt{uci\_concrete} this is an ordinary sampling outcome: both intervals straddle zero (Table~\ref{tab:8c}). \texttt{cross\_fitted\_normalized} on \texttt{uci\_superconductivity} does not fit that explanation, since its interval of [-0.011, -0.005] excludes zero, so the null over-predicts the worst-bin shortfall for that cell rather than merely being met. We note it without resolving it; \texttt{split\_conformal} on \texttt{uci\_concrete}
(-0.004, Table~\ref{tab:8}) is a third such case.

\subsection{Dataset-specific observations}

On \texttt{matbench\_expt\_gap}, 53\% of target values were exactly 0.0. The lower half
of the quartile partition was degenerate under this condition. CQR's
correction term Q collapsed to 0.0 on all 50 seeds on this dataset, making
its intervals bit-identical to raw quantile regression's uncorrected
intervals; both methods' excess and significance-rate values on
\texttt{matbench\_expt\_gap} in Tables~\ref{tab:3}--\ref{tab:9} are therefore equal by construction rather
than by independent measurement.

On \texttt{uci\_superconductivity}, \texttt{mondrian}'s on-partition significance rate was 0.24
(quartile) and 1.00 (octile); this was the only dataset on which \texttt{mondrian}
exceeded the octile null's 95th percentile on all 50 seeds.

\subsection{Interval width}

Table~\ref{tab:10} reports mean interval width by prediction quartile for SCP and
Mondrian, averaged across 50 seeds.

\begin{table}[htbp]
\centering
\footnotesize
\caption{Mean interval width by quartile}
\label{tab:10}
\begin{tabular}{llrrrr}
\toprule
Dataset & Method & Q1 & Q2 & Q3 & Q4 \\
\midrule
\texttt{california\_housing} & \texttt{mondrian} & 0.832 & 1.207 & 1.660 & 2.180 \\
\texttt{california\_housing} & \texttt{split\_conformal} & 1.537 & 1.537 & 1.537 & 1.537 \\
\texttt{matbench\_expt\_gap} & \texttt{mondrian} & 0.228 & 1.248 & 2.940 & 4.329 \\
\texttt{matbench\_expt\_gap} & \texttt{split\_conformal} & 2.686 & 2.686 & 2.686 & 2.686 \\
\texttt{nims\_matnavi} & \texttt{mondrian} & 71.08 & 65.76 & 76.64 & 192.15 \\
\texttt{nims\_matnavi} & \texttt{mondrian\_matched} & 54.85 & 49.00 & 56.91 & 151.59 \\
\texttt{nims\_matnavi} & \texttt{split\_conformal} & 93.90 & 93.90 & 93.90 & 93.90 \\
\texttt{uci\_concrete} & \texttt{mondrian} & 11.91 & 16.45 & 16.74 & 22.01 \\
\texttt{uci\_concrete} & \texttt{mondrian\_matched} & 10.66 & 14.85 & 15.15 & 19.77 \\
\texttt{uci\_concrete} & \texttt{split\_conformal} & 16.35 & 16.35 & 16.35 & 16.35 \\
\texttt{uci\_superconductivity} & \texttt{mondrian} & 9.16 & 21.01 & 53.88 & 46.92 \\
\texttt{uci\_superconductivity} & \texttt{split\_conformal} & 37.17 & 37.17 & 37.17 & 37.17 \\
\bottomrule
\end{tabular}
\end{table}

SCP's width is identical across quartiles on every dataset (Q4/Q1 = 1.000),
as follows from its construction: a single pooled quantile of the
calibration residuals is applied to every test point.

Mondrian's width varied across quartiles on all five datasets. The variation
was monotonically increasing on \texttt{california\_housing}, \texttt{matbench\_expt\_gap}, and
\texttt{uci\_concrete}. It was non-monotonic on \texttt{nims\_matnavi} (Q2 narrower than Q1) and
on \texttt{uci\_superconductivity}, where the widest quartile was Q3 (53.88) rather
than Q4 (46.92).

Averaged across quartiles, Mondrian's mean width was below SCP's on
\texttt{california\_housing} (1.470 vs. 1.537), \texttt{matbench\_expt\_gap} (2.186 vs. 2.686) and
\texttt{uci\_superconductivity} (32.74 vs. 37.17), and above it on \texttt{uci\_concrete} (16.78
vs. 16.35) and \texttt{nims\_matnavi} (101.4 vs. 93.90).

The pooled comparison conceals the size of the redistribution. On
\texttt{uci\_concrete}, where Mondrian's mean width exceeded SCP's by 2.6\%, Mondrian's
Q1 intervals were 27\% narrower than SCP's and its Q4 intervals 35\% wider. On
\texttt{nims\_matnavi}, where Mondrian's mean width exceeded SCP's by 8.0\%, its Q1--Q3
intervals were narrower than SCP's by 24\%, 30\% and 18\% respectively, while
its Q4 intervals were 105\% wider.

The level-matched variant, run on \texttt{nims\_matnavi} and \texttt{uci\_concrete}, was narrower
than standard Mondrian by a near-constant factor within each dataset: the
matched-to-standard width ratio was 0.77, 0.75, 0.74 and 0.79 across Q1--Q4 on
\texttt{nims\_matnavi}, and 0.90 in all four quartiles on \texttt{uci\_concrete}. The shape of the
width profile was preserved. The Q4-to-Q1 width ratio was 2.70 for standard
Mondrian and 2.76 for the matched variant on \texttt{nims\_matnavi}, and 1.85 against 1.86 on \texttt{uci\_concrete}; width ratios throughout are computed from unrounded widths and so need not reproduce exactly from the tabulated two-decimal values. Averaged across quartiles, the matched variant's mean
width was 78.09 on \texttt{nims\_matnavi} and 15.11 on \texttt{uci\_concrete}, in both cases below
the corresponding SCP width (93.90 and 16.35).

\section{Discussion}

\subsection{What the null correction changes}

The min-over-bins statistic used in most conditional-coverage evaluations,
including that of \citet{boruah2026}, is biased downward relative to the
target it is meant to
test. This is not a new observation: the statistic is worst-slab coverage
\citep{romano2020,cauchois2021}, its finite-sample negative bias
is recognised, and implementations mitigate it by splitting the evaluation
sample, while \citet{braun2025} avoid the binned statistic altogether by
recasting the diagnostic as classification. What is new here is not the
diagnosis but its absence from applied materials UQ, and the size of the
consequence when it is supplied. Concretely: even a method with exact conditional validity will show an apparent
deficit, because the minimum of several noisy per-bin estimates falls below
the true rate by construction. Table~\ref{tab:3} shows this bias is not a minor
correction. On \texttt{nims\_matnavi}, an exactly valid method is expected to show a
quartile min-gap of 0.061 to 0.096 from finite-sample noise alone (Results,
Table~\ref{tab:3}), before any real non-uniformity is added. The equivalent figure on
the two largest datasets is 0.009 to 0.011.
On \texttt{nims\_matnavi}, three of the seven methods, \texttt{cqr}, \texttt{mondrian}, and \texttt{mondrian\_matched},
have raw quartile min-gaps that owe most of their size to this
effect rather than to a genuine coverage failure.

This matters beyond bookkeeping. A method's reported "worst-quartile
coverage" is routinely treated as a direct estimate of how badly it fails
conditionally. Under that reading, \texttt{cqr}'s 0.816 on \texttt{nims\_matnavi} in
\citet{boruah2026} and
\texttt{mondrian\_matched}'s 0.865 marginal coverage in this paper's own ablation would
both look like meaningful failures. Once the null is subtracted, most of both is accounted for: \texttt{cqr}'s quartile excess on \texttt{nims\_matnavi} falls to 0.061, roughly half the raw shortfall, and exceeds the null's 95th percentile on 24\% of seeds --- well below the majority criterion used here, though still above the 0.05 rate expected under exact conditional validity, so a small real component cannot be ruled out. \texttt{mondrian\_matched}'s apparent
quartile gap is explained by its 0.865 marginal coverage rather than by
uneven distribution of that coverage across bins (Discussion, "Level and
non-uniformity are separable," below).

\subsection{The antecedent study's negative case, reconsidered}

\citet{boruah2026} identified conformalized quantile regression as a negative case:
marginally valid at 0.920 but reaching only 0.816 in the top quartile against
a 0.90 target, with a quartile range of 0.172 and the widest intervals of any
method in every quartile. It attributed this to a single additive correction
being applied uniformly to quantile estimates miscalibrated by differing
amounts in different regions, and scoped the claim explicitly to n = 437 with
those base learners rather than to CQR in general. The null-corrected analysis
does not contradict the 0.816. It changes what can be inferred from it. \texttt{cqr}'s
quartile excess on \texttt{nims\_matnavi} (0.061) exceeds the null's 95th percentile on 24\% of seeds, against 0.05 under exact conditional validity and against the majority criterion used here. Across the
five-dataset replication, \texttt{cqr} exceeded the null on exactly one dataset,
\texttt{matbench\_expt\_gap}, where its correction term collapsed to zero on all 50
seeds and its intervals became bit-identical to raw quantile regression's
(Results, "Dataset-specific observations"). On every other dataset, including
the feature partition throughout, \texttt{cqr}'s excess was among the lowest of the
seven methods studied.

The antecedent study already declined to generalise from that dataset. What
the replication changes is the attribution rather than the scope. It read the Q4 shortfall as evidence that a global additive offset cannot repair
locally varying quantile miscalibration. Against the null, roughly half of \texttt{cqr}'s on-partition shortfall on \texttt{nims\_matnavi} is expected under exact conditional validity, and what remains reaches the 95th percentile on a quarter of seeds --- too little to carry a mechanism, though not nothing. The one dataset where \texttt{cqr} does exceed the null on a majority of seeds, \texttt{matbench\_expt\_gap}, fails for an unrelated reason: the correction term collapses to zero on all 50 seeds and
the intervals become bit-identical to uncorrected quantile regression's
(Results, "Dataset-specific observations"). That is a property of a target
distribution with 53\% of its mass at a single point, not of the
conformalization step. The 0.816 stands as reported; the mechanism proposed
for it there does not survive the correction, and \texttt{cqr}'s
excess is among the lowest of the seven methods on every other dataset and on
the feature partition throughout.

\subsection{The three-tier structure does not hold as stated}

\citet{boruah2026} grouped cross-fitted normalized conformal prediction (NormCP-CF)
and Mondrian together as locally-adaptive methods that succeed conditionally, distinct from constant-width methods (SCP, GPR) that fail. The
null-corrected results separate these two methods. NormCP-CF's quartile
excess exceeded the null's 95th percentile on 100\% of seeds on both
\texttt{california\_housing} and \texttt{uci\_superconductivity}, and on 82\% on
\texttt{matbench\_expt\_gap}, against \texttt{split\_conformal}'s 1.00 on all three. (The comparison is drawn against \texttt{split\_conformal} rather than \texttt{gpr}
because \texttt{gpr} was subsampled on all three, as set out under Limitations.)
Mondrian reached the majority criterion on the quartile partition on none of the five datasets, with exceedance rates of 0.10 to 0.24. Those rates sit two to five times the 0.05 expected under exact conditional validity, so the claim is that Mondrian's on-partition deficit is small rather than that it is absent.

Counting datasets on which each method's quartile excess exceeded the null's
95th-percentile threshold on more than half of seeds: \texttt{split\_conformal} (4),
\texttt{gpr} (3), \texttt{cross\_fitted\_normalized} (3), \texttt{raw\_quantile} (2), \texttt{cqr} (1), \texttt{mondrian} (0), \texttt{mondrian\_matched} (0). Two entries should not be read as conditional performance: \texttt{raw\_quantile} is marginally invalid at 0.770 and 0.782, so its null is built at a lower reference and its conditional test is correspondingly easier, and \texttt{gpr}'s three datasets are exactly the three on which it was subsampled (Limitations). Among the four methods that are marginally valid and saw the full training split, the ordering is \texttt{split\_conformal} (4), \texttt{cross\_fitted\_normalized} (3), \texttt{cqr} (1), \texttt{mondrian} (0). This does not support a two-tier split between
"constant-width, failing" and "locally-adaptive, succeeding" methods.
NormCP-CF's local adaptivity did not prevent conditional failure on three of
five datasets. The distinction that holds is one of degree rather than of kind: \texttt{mondrian}'s on-partition exceedance rate stayed within 0.10 to 0.24 on every dataset, while every other method studied reached 0.60 or above on at least one. The interval estimates make the same point without reference to any threshold: on \texttt{california\_housing} \texttt{mondrian}'s quartile excess is 0.003 [0.001, 0.005] against \texttt{split\_conformal}'s 0.098 [0.095, 0.101], and the two intervals are separated by more than an order of magnitude on all three of the largest datasets (Table~\ref{tab:4c}). That separation is the result the paper rests on. It is a difference in magnitude, not a boundary between failing and not failing, and every method here including \texttt{mondrian} has a quartile excess whose interval excludes zero.

\subsection{Level and non-uniformity are separable}

The antecedent study raised this question itself, noting that roughly 22 calibration
points per group put Mondrian's finite-sample level at 0.954 against the
pooled 0.920, and asserting that the uniformity of its coverage was a real
property of the construction while the level at which that uniformity sat was
partly an artifact of sample size. That assertion was not tested there. The
ablation below supplies the test.

That the finite-sample correction inflates the effective level as per-group
calibration sets shrink has also been observed independently:
\citet{rafe2026} make the same point for Mondrian calibration on thin
demographic cells in survey data. That study reaches the opposite empirical conclusion to ours,
finding that Mondrian increases rather than decreases the worst-subgroup
coverage gap, and proposes shrinking group thresholds toward the global
quantile as a remedy. The divergence is worth stating plainly rather than
eliding. Their groups are externally defined demographic cells that need not
align with residual scale, whereas ours are quantiles of the predicted value
and therefore align with heteroscedasticity by construction; the level-matched
ablation below indicates the mechanism is the group-conditional calibration
itself rather than the elevated level, which suggests the difference lies in
whether the grouping variable tracks residual scale. We do not have their data
and cannot test that directly.

The effective-level analysis (Results, Table~\ref{tab:2}) showed that Mondrian's
finite-sample correction inflates its per-bin quantile level well above
nominal on small datasets: a mean level of 0.965 across bins on
\texttt{nims\_matnavi}, against a target of 0.90, with individual bins reaching 1.0 on
41 of 50 seeds. A natural question is whether Mondrian's apparent
conditional success is partly an artifact of this inflation: a method
targeting a higher level would show flatter coverage even without any real
conditional structure, simply because it is more conservative everywhere.

The level-matched ablation addresses this directly. \texttt{mondrian\_matched} targets
1-$\alpha$ exactly, with no finite-sample correction, and its marginal coverage
(0.865 on \texttt{nims\_matnavi}, 0.882 on \texttt{uci\_concrete}) confirms the level was in fact
reduced to the intended target. Its quartile coverage profile remained flat
(Results, "Effective Mondrian level"), and its null-corrected quartile excess
after conditioning on its own observed marginal coverage (0.039 on
\texttt{nims\_matnavi}, 0.020 on \texttt{uci\_concrete}) was close to plain Mondrian's (0.033,
0.021) rather than close to zero. The same held off-partition: octile excess
of 0.045 and 0.031 for the matched variant against 0.029 and 0.027 for
standard Mondrian. Level-matching moved marginal coverage without materially
moving the excess statistic on any partition, on- or off-. Mondrian's
within-partition uniformity is a property of the group-conditional
calibration step, not a byproduct of the finite-sample correction operating
at an elevated level.

The interval-width results (Results, Table~\ref{tab:10}) provide independent support.
Mondrian's width varies across quartiles on all five datasets while SCP's is
constant by construction, and the direction of that variation tracks residual
scale rather than predicted magnitude. On four datasets Mondrian's widest
quartile is Q4. On \texttt{uci\_superconductivity} it is Q3 (53.88, against 46.92 in
Q4), and \texttt{uci\_superconductivity} is the one dataset in the study where SCP's
coverage failure also falls in Q3 (0.778) rather than Q4 (0.841). A method
widening intervals in proportion to predicted magnitude would place its widest
intervals in Q4 on every dataset; Mondrian instead widens where the residual
scale actually is.

The same results caution against reading pooled mean width as the cost of
group-conditional calibration. On \texttt{uci\_concrete}, Mondrian's mean width exceeds
SCP's by 2.6\%, a difference that would suggest near-equivalence; the
underlying per-quartile intervals range from 27\% narrower than SCP in Q1 to
35\% wider in Q4. On \texttt{nims\_matnavi} the pooled excess is 8.0\%, while Q1 through
Q3 are 24\%, 30\% and 18\% narrower than SCP and only Q4 is wider, by 105\%. The
aggregate figure conceals a redistribution substantially larger than itself in
both cases.

The level-matched ablation separates the two effects that produce this
profile. Removing the finite-sample correction shrinks Mondrian's intervals by
a near-constant factor within each dataset. The ratio of matched to standard
width is 0.77, 0.75, 0.74 and 0.79 across Q1 to Q4 on \texttt{nims\_matnavi}, and 0.90
in every quartile on \texttt{uci\_concrete}, leaving the shape of the width profile
essentially unchanged. The Q4-to-Q1 width ratio is 2.70 for standard Mondrian
and 2.76 for the matched variant on \texttt{nims\_matnavi}, and 1.85 against 1.86 on
\texttt{uci\_concrete}.

Two consequences follow. First, Mondrian's pooled width excess over SCP is
attributable to the elevated effective level rather than to group-conditional
calibration: at a matched level, Mondrian's mean width is 78.09 on \texttt{nims\_matnavi}
against SCP's 93.90, and 15.11 on \texttt{uci\_concrete} against 16.35, below SCP on
both datasets, where the uncorrected comparison places it above on both.
Second, the concentration of width in Q4 is not an artifact of quantile
saturation. Saturation cannot occur under the matched variant by construction,
yet its Q4 intervals remain 2.76 times its Q1 intervals and remain wider than
SCP's uniform interval. The width profile tracks residual scale, consistent
with the 2.7-fold elevation of Q4 residual spread reported for this dataset in
the antecedent study.

\subsection{The guarantee holds at its own resolution and not beyond it}

The relationship between Mondrian's on-partition and octile performance on the
same underlying variable is the clearest result in this replication, though
the phenomenon it exhibits is not new. Calibrating on a coarse partition and
evaluating on a refinement of it is the central design of clustered conformal
prediction \citep{ding2023}, where classes are pooled into clusters,
validity is guaranteed at the cluster level, and the resulting per-class
coverage gap is shown to depend on how similar the pooled classes are. Our
setting differs in that the refinement is a quantile refinement of a
continuous predicted variable rather than a set of discrete labels within
clusters, and in that the comparison is null-corrected; the underlying
phenomenon is the same one. The mechanism is not mysterious, and stating it plainly is worth more than presenting the result as a discovery. Mondrian applies a single pooled quantile within each calibration group, so within a group it is split conformal prediction; residual scale that varies inside a quartile must therefore produce an octile coverage gradient by exactly the mechanism that produces split conformal prediction's quartile gradient. What the replication contributes is not that the effect exists but how large it is, and how that size depends on test-set size. The octile partition is not an independent test of Mondrian; it is a refinement of the exact partition Mondrian calibrates on, built from the same out-of-fold training predictions (Results, "A null-corrected statistic
for min-over-bins coverage"). On the calibration partition itself, Mondrian's quartile excess reached the null's 95th percentile on 0.10 to 0.24 of seeds depending on the dataset, peaking at 0.24 on superconductivity, and so stayed below the majority criterion on all five. That peak is above the 0.05
rate expected under exact conditional validity, so the on-partition claim is
not that Mondrian is exactly conditionally valid at quartile resolution. On
the octile refinement of that same partition, the rate reached 0.66--1.00 on
three of five datasets (\texttt{california\_housing}, \texttt{matbench\_expt\_gap},
superconductivity). The contrast that carries the argument is internal to
Mondrian and independent of where the significance line is drawn: the same
method, calibrated on the same variable, moves from 0.24 to 1.00 on
superconductivity when the evaluation partition is refined from four bins to
eight.

Mondrian's group-conditional guarantee is defined with respect to the
partition it calibrates on; it was never a guarantee at finer resolution, and
the octile result should not be read as a violation of a promise the method
made. It does bound how far a practitioner should expect that guarantee to
transfer to groupings finer than the one specified. On the two smallest
datasets in the replication, \texttt{nims\_matnavi} and \texttt{uci\_concrete}, the octile
partition failed to reach significance on either dataset. The smallest
octile bins on \texttt{nims\_matnavi} held as few as one test point in a given seed,
so this is better read as insufficient statistical power than as evidence
that Mondrian's guarantee extends to octile resolution on small data. The three
datasets large enough to test the question all showed the same pattern:
exact at the calibration resolution, degraded at twice that resolution.

The feature partition, which conditions on a variable orthogonal to the one
Mondrian calibrates on rather than a refinement of it, showed lower excess than the octile partition in 31 of the 32 method-dataset cells (Results, Table~\ref{tab:8} vs. Table~\ref{tab:6}). The exception is \texttt{mondrian} on \texttt{nims\_matnavi}, where the feature excess of 0.036 exceeds the octile excess of 0.029; the two are within sampling error of each other (Table~\ref{tab:6c} and Table~\ref{tab:8c}), and \texttt{nims\_matnavi} is the dataset whose octile bins are too small to resolve a deficit of this size at all. This is consistent with the octile result being driven
specifically by within-partition heterogeneity along the calibrated
variable, rather than by conditional coverage failure in general.

\subsection{Limitations}

The octile partition's power to detect failure depends on dataset size in a
way that limits interpretation on the two smallest datasets studied. Neither
\texttt{nims\_matnavi} nor \texttt{uci\_concrete} reached statistical significance on the octile
partition for any method, but both have octile bins small enough (minimum 1
and 11 test points respectively, across seeds) that a real deficit of the
size seen on the larger datasets would likely go undetected. The absence of
a significant result there should not be read as evidence that Mondrian's
guarantee holds at octile resolution in general; it is evidence only for
these two datasets at this sample size.

GPR's conditional failures cannot be separated from its training-set
subsampling. Cubic scaling in sample size required capping its training split
at 2000 points on the three largest datasets, and those are exactly the three
on which it exceeded the null on a majority of seeds; on the two datasets
where it saw the full training split its significance rates were 0.38 and
0.16. A second artifact acts in the same direction, since larger test sets
give more power to detect a deficit of given size. The two cannot be
disentangled within this design, and \texttt{gpr}'s position in the ordering of Table~\ref{tab:5}
should not be read as a property of Gaussian process regression.

The feature-partition selection rule excludes features with fewer than eight
distinct values or non-strictly-increasing quartile cut points, which is
necessary to avoid degenerate bins on composition data but means the
partition tested is not necessarily the most diagnostic one available for
each dataset. A different selection rule could show a different pattern.

The null-corrected statistic was evaluated at $\alpha$ = 0.10 throughout, matching
\citet{boruah2026}. Whether the same qualitative pattern, on-partition success and
resolution-dependent off-partition degradation, holds at other significance
levels was not tested.

All seven methods are built on a single point model, a gradient boosting
regressor with hyperparameters frozen at the antecedent study's values across
every dataset. Freezing it keeps the interval-construction comparison clean,
since no column is advantaged by a better-tuned mean function. The cost is
that every result here describes conformalization applied to one particular
regressor. Residual structure that drives conditional coverage failure is a
joint property of model and data, so a different base learner, or the same one
tuned per dataset, could redistribute residual scale across the prediction
range and change which methods fail where.

The significance rates reported here are per-cell and uncorrected for
multiplicity. Six methods on five datasets and a seventh on two, each evaluated on three partitions, give 96 method-dataset-partition combinations, so some cells near the criterion would
be expected to cross it by chance. This bears on marginal cases rather than on
the results the paper rests on: \texttt{split\_conformal}'s rates of 1.00 on three
datasets and Mondrian's move from 0.24 to 1.00 under partition refinement are
not sensitive to any multiplicity adjustment, whereas single borderline cells
such as \texttt{cross\_fitted\_normalized} at 0.82 on \texttt{matbench\_expt\_gap} should be read
with the count of comparisons in mind.

\section{Conclusion}

Conditional coverage in materials property prediction is routinely reported as
the minimum coverage across strata of the predicted value. That statistic is
downward-biased in finite samples, and this paper measures how much of the reported failure it accounts for. Across five regression datasets spanning
437 to 21,263 samples, seven interval methods and 50 splits each, the expected
shortfall under exact conditional validity ranges from 0.009 on the largest
datasets to 0.096 on the smallest. On a 437-sample fatigue dataset it consumes
between a third and two thirds of the shortfalls actually reported.

Split conformal prediction's conditional failure survives the correction on
four of five datasets. Two conclusions of the antecedent single-dataset study
do not. Conformalized quantile regression, characterized there as a negative
case, exceeds the null on one dataset only, and on that dataset its correction
term degenerates to zero and its intervals coincide with uncorrected quantile
regression's. Cross-fitted normalized conformal prediction, grouped there with
the methods that succeed, fails on three of five. The three-tier grouping of
methods by local adaptivity does not survive the corrected statistic.

Group-conditional calibration behaves as its guarantee specifies and not
beyond it. Mondrian conformal prediction is the only method of the seven whose on-partition exceedance rate stays below the majority criterion on every dataset, at 0.10 to 0.24 against 0.60 or more for every other method on at least one dataset, and it exceeds the null on three of five
when evaluated on an octile refinement of the same variable it calibrates on.
The level-matched ablation locates this in the group-conditional calibration
step rather than in the elevated effective quantile level that the
finite-sample correction produces on small per-group calibration sets.
Removing the correction shifts marginal coverage to 0.865 and 0.882, and
leaves both the excess statistic and the shape of the width profile
essentially unchanged. It also removes Mondrian's apparent width cost against
split conformal prediction entirely.

None of the three corrections applied here is new to the conformal prediction
literature. They have simply not been applied in this domain, on datasets
small enough for the bias to dominate what gets reported. Conditional coverage
evaluations here should report the null distribution of whatever worst-bin
statistic they use, the method's realized marginal coverage, and the
resolution of the evaluation partition relative to any partition the method
calibrates on.

\section*{Data and code availability}

The analysis code, evaluation harness and per-seed result files supporting this
study are archived at Zenodo, https://doi.org/10.5281/zenodo.22213087, released
under the MIT licence. All five datasets are public: \texttt{nims\_matnavi} from the NIMS
MatNavi fatigue database, available from Japan's National Institute for
Materials Science subject to their access terms \citep{furuya2019,nims2026};
\texttt{uci\_concrete} and \texttt{uci\_superconductivity} from the UCI Machine Learning Repository
\citep{yeh1998data,hamidieh2018data}; \texttt{matbench\_expt\_gap} via \texttt{matminer}
\citep{dunn2020,ward2018}; and \texttt{california\_housing} via scikit-learn
\citep{pace1997,pedregosa2011}. Featurization, splits and seeds are fully
specified in Methods, and the deposited harness reproduces every table from raw
data.

\section*{Declarations}

\textbf{Funding statement.} This research received no specific grant from any
funding agency.

\textbf{Competing interest.} None.

\textbf{Author contribution.} The author conceived the study, implemented the
evaluation harness, performed the computational analysis, and wrote the
manuscript.

\textbf{AI use statement.} Generative AI tools were used to assist with code
debugging, LaTeX troubleshooting, and prose editing during manuscript
preparation. All experimental design, analysis, interpretation, and conclusions
are the author's own.

\textbf{Ethical standard.} The research meets all ethical guidelines, including
adherence to the legal requirements of the study country.

\textbf{Related work by the author.} This paper extends and partly revises
\citet{boruah2026}, currently under review at this journal (manuscript ID
DCE-2026-0115). The two are submitted separately rather than merged because the five-dataset scope, the null construction and the level-matched ablation are a distinct contribution, and because the antecedent study's reported numbers stand as measured: what changes is what may be inferred from them. The editor has been notified of the relationship between the two submissions.

\printbibliography

\end{document}